\documentclass[twocolumn,notitlepage,nofootinbib]{revtex4-2}
\usepackage{amsmath,amssymb}
\usepackage{graphicx}
\usepackage{newtxtext}
\usepackage{newtxmath}
\usepackage[x11names]{xcolor}
\usepackage[unicode,colorlinks,citecolor=Blue3,linkcolor=Blue3,bookmarks=true]{hyperref}

\newcommand\bra[2][]{#1\langle {#2} #1|}
\newcommand\ket[2][]{#1|{#2} #1\rangle}
\newcommand{\braket}[2]{ \langle #1 | #2 \rangle}
\renewcommand{\Re}{\mathop{\rm Re}\nolimits}

\begin{document}

\title{Using Schr\"odinger cat quantum state for detection of a given phase shift}

\author{V.\,L.\,Gorshenin}
\email{valentine.gorshenin@yandex.ru}
\affiliation{Independent Researcher, Moscow, Russia}

\begin{abstract}

We show that injecting a light pulse prepared in the Shr\"odinger cat quantum state into
the dark port of a two-arm interferometer, it is possible to detect a given phase shift unambiguously. The value of this phase shift is inversely proportional to the amplitudes of both the classical carrier light and the Shr\"odinger cat state. However, an unconventional detection procedure is required for this purpose.

By measuring the number of photons at the output dark port, it is possible to detect the phase shift with a vanishing ``false positive'' probability. The ``false negative''  probability in this case decreases as the amplitude of the Schr\"odinger cat state increases and, for reasonable values of this amplitude, can be made less than about 0.1.

\end{abstract}

\maketitle
\section{Introduction}

On the fundamental level, the phase sensitivity of the interferometers is limited by quantum fluctuations of the probing light and, therefore, depends on its quantum state, see e.g. the review papers \cite{Anderson_ch35_2019, 22a1SaKh}. In the most basic case of coherent quantum states, generated by phase-stabilized lasers, the phase sensitivity corresponds to the shot noise limit (SNL):
\begin{equation}\label{SNL}
	\Delta\phi_{\rm SNL} = \frac{1}{2\sqrt{N}} \,,
\end{equation}
where \(N\) is the number of photons used for measurement (that is the ones that interacted with the phase shifting object(s)).

Better sensitivity, for a given value of \(N\), can be achieved by using squeezed quantum states of light \cite{Caves1981}. In the case of moderate squeezing, $e^{2r}\ll N$, the phase sensitivity could be improved by factor $e^r$ in comparison with SNL:
\begin{equation}\label{dphi_sqz}
	\Delta\phi_{\rm SQZ} = \frac{e^{-r}}{2\sqrt{N}} \,,
\end{equation}
where \(r\) is the logarithmic squeeze factor. This method is successfully used in the  kilometer scale interferometers of the modern gravitational-waves (GW) detectors  \cite{Dwyer_Galaxies10_020046_2022}.

In case of very strong squeezing, $e^{2r}\gtrsim N$, the phase sensitivity is limited by Heisenberg Limit (HL) \cite{Ou_PRL_77_2352_1996, Ou_PRA_55_2598_1997, 17a1MaKhCh}:
\begin{equation}\label{HL}
	\Delta\phi_{\rm HL} \sim \frac{1}{N} \,.
\end{equation}

Both coherent and squeezed states belong to the class of the Gaussian states: their Wigner quasi-probability functions \cite{Schleich2001} have a Gaussian form. The use of more sophisticated non-Gaussian quantum states has also been considered in the literature, see {\it e.g.} the articles \cite{Holland_PRL_71_1355_1993, Lee_JMO_49_2325_2002, Campos_PRA_68_023810_2003, Berry_PRA_80_052114_2009, Pezze_PRL_110_163604_2013, Daryanoosh_NComm_9_4606_2018, Perarnau-Llobet_QST_5_025003_2020, Shukla_OptQEl_55-460_2023, Shukla_PhOpen_18_100200_2024}. In particular, in Refs.\,\cite{Shukla_OptQEl_55-460_2023, Shukla_PhOpen_18_100200_2024}, the use of ``Shr\"odinger cat'' (SC) quantum states of the form \eqref{std_cat} was explored theoretically in this context. However, as it was shown in Refs.\,\cite{Lang_PRL_111_173601_2013, Lang_PRA_90_025802_2014}, the optimal sensitivity could be provided by less exotic and easier to prepare Gaussian states.

In these works, the problem of measuring of an {\it a priori} unknown phase was considered, with \eqref{SNL}-\eqref{HL} being the mean square error of this measurement. Another standard problem of the detection and estimation theory is the discrimination of two possible hypotheses \cite{HelstromBook}, and, in particular, the binary (yes/no) {\it detection} of a given phase shift. Potentially, this approach could provide better sensitivity, thanks to the {\it a priori} information on the signal. It can be used, for example, for the discrimination of samples with two slightly different, but known in advance, values of the refractive indices.

In this case, non-Gaussian quantum states could provide significantly better detection fidelity than the Gaussian ones because they can be orthogonal to each other and thus can be discriminated unambiguously \cite{HelstromBook}. This concept was experimentally demonstrated in Ref.\,\cite{Wolf_NComm_10_2929_2019} for the detection of an external force acting on an ion in the trap, with its translational degree of freedom  prepared in the non-Gaussian Fock state.

In Ref.\,\cite{Singh_PhOpen_18_100198_2024} it was shown that the quantum state, prepared by applying the unitary displacement operator $\mathcal{D}$ to the SC state, can be orthogonal to the initial one for certain values of the displacement parameter. Therefore, SC states can be used for the unambiguous detection of this displacement. However, no specific measurement procedure was considered in that paper.

In our work, we consider the optical interferometric schemes (see Fig.\,\ref{fig:parity-optic-scheme}) that use the SC state for detecting a given phase shift. The paper is organized as follows. In section \ref{sec:ifos} we show that in the linearized case of a strong classical carrier and small phase shift, the evolution of light in the standard two-arm interferometer can be described as the action of the displacement operator $\mathcal{D}$. In section \ref{sec:orthog}, we calculate  the phase shift that provides the orthogonality of the initial and displaced SC states. In section \ref{sec:detection} we calculate the sensitivity, assuming the photon number measurement at the output dark port. Finally, in section \ref{sec:conclusion}, we summarize our results and discuss the potential practical implementation of the proposed scheme.

\section{Evolution of light in the interferometer}\label{sec:ifos}

\begin{figure}
	\centering
	\includegraphics[width=1\linewidth]{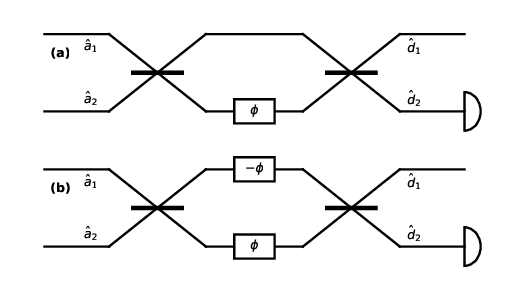}
	\caption{Optical schemes of the asymmetric (a) and antisymmetric (b) Mach–Zehnder interferometers. The bright ports denoted by the subscript ``1'', and the dark ones --- by ``2''.}
	\label{fig:parity-optic-scheme}
\end{figure}

Following the review \cite{22a1SaKh}, we consider two practically important configurations of the Mach–Zehnder interferometer --- the asymmetric one, see Fig.\,\ref{fig:parity-optic-scheme}(a), and the antisymmetric one, see Fig.\,\ref{fig:parity-optic-scheme}(b). In the conceptually simpler asymmetric case,  the signal phase shift is introduced in the first arm, with the second one providing the reference beam. The amplitude reflectivity $R$ and the transmissivity $T$ of the beamsplitters in this case could differ from each other, $R\ne T$. In the second configuration, the phase shift is introduced antisymmetrically in both arms, and the balanced beamsplitters have to be used, $R=T=1/\sqrt{2}$. This variant is not sensitive to the common phase shift and therefore more tolerant to technical noises and drifts. Due to this reason, the antisymmetric configuration is used, in particular, in the GW detectors \cite{CQG_32_7_074001_2015} (strictly speaking, GW detectors use the Michelson  interferometer topology; it is well known, however, that Michelson and Mach-Zehnder topologies are equivalent to each other).

In both cases, we assume that the strong coherent carrier light is fed into the bright input port (the first one in Figs.\,\ref{fig:parity-optic-scheme}), and some quantum state is injected into the dark input port (the second one in Figs.\,\ref{fig:parity-optic-scheme}). We also assume that the interferometer is tuned in such a way that in the absence of the phase signal ($\phi=0$), both input states are reproduced at the respective bright and dark outputs.

The input/output relations for the schemes of Figs.\,\ref{fig:parity-optic-scheme} are calculated in App.\,\ref{app:ifo} using the Heisenberg picture. It is shown that in the  linear in small phase shift $\phi$ and quantum fluctuations approximation, the optical field at the bright output port does not depend on $\phi$, while the optical field at dark output port can be presented as follows:
\begin{equation}\label{ifo_io_H}
	\hat{d}_2 = \hat{a}_2 + iB\phi \,,
\end{equation}
where $\hat{a}_2$, $\hat{d}_2$ are the annihilation operators at the dark input and output ports, respectively, and
\begin{equation}
	B=\sqrt{N}
\end{equation}
In the asymmetric case, $B=TA$, while in the antisymmetric one, $B=A$. In both cases $A>0$ is the classical carrier amplitude at the interferometer input.

Equation \eqref{ifo_io_H} can be rewritten as follows:
\begin{equation}
	\hat{d}_2 = \hat{\mathcal{D}}^\dag(\delta)\hat{a}_2\hat{\mathcal{D}}(\delta) \,,
\end{equation}
where
\begin{equation}\label{D_delta}
	\hat{\mathcal{D}}(\delta) = e^{i\delta(\hat{a}_2^\dag + \hat{a}_2)} \,,
\end{equation}
is the displacement operator and
\begin{equation}\label{delta}
	\delta = B\phi \,.
\end{equation}
In the Schr\"odinger picture, Eq.\,\eqref{ifo_io_H} translates to the following relation:
\begin{equation}\label{ifo_io_S}
  \ket{\Psi_\delta} = \hat{\mathcal{D}}(\delta)\ket{\Psi_0} \,,
\end{equation}
where $\ket{\Psi_0}$ and $\ket{\Psi_\delta}$ are the quantum states of light at, respectively, the input and output dark ports.

\section{The potential sensitivity}\label{sec:orthog}

Suppose that the incident light at the dark port is prepared in the SC state:
\begin{equation}\label{std_cat}
	\ket{\Psi_0} = \frac{1}{\sqrt{K}}(\ket{\alpha} + \ket{-\alpha}) \,,
\end{equation}
where $\ket{\alpha}$ and $\ket{-\alpha}$ are the coherent states and
\begin{equation}
	K = 2(1+e^{-2|\alpha|^2})
\end{equation}
is the normalization factor.

\begin{figure}
	\centering
	\includegraphics[width=1\linewidth]{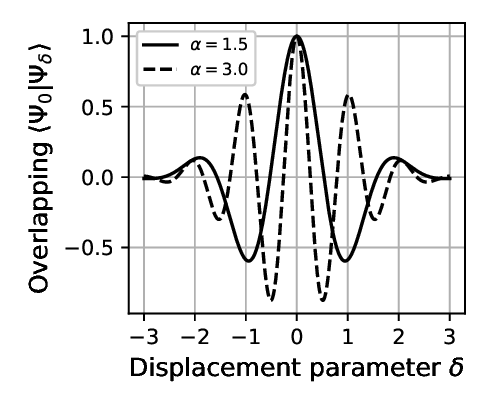}
	\caption{Overlapping of the initial and displaced SC states (see Eq.\,\eqref{overlap}) as a function of the displacement parameter $\delta$  for \(\alpha = 1.5\) (solid line) and \(\alpha = 3\) (dashed line).}
	\label{fig:plot-overlap-by-phi-fixed-alpha-2}
\end{figure}

It can be shown that in order to obtain the best sensitivity with a displacement operator of the form \eqref{D_delta}, with the real displacement parameter $\delta$, the parameter $\alpha$ also has to be real. This corresponds to displacement of the SC state in the direction orthogonal to SC interference strips. In this case, the output quantum state is equal to
\begin{equation}\label{Psi_delta}
  \ket{\Psi_\delta} = \frac{1}{\sqrt{K}}
	 (e^{i\delta\alpha}\ket{\alpha+i\delta} + e^{-i\delta\alpha}\ket{-\alpha+i\delta}) \,.
\end{equation}
Taking into account that for any complex numbers $\alpha_{1,2}$,
\begin{equation}
	\braket{\alpha_1}{\alpha_2} = e^{-(|\alpha_1|^2 + |\alpha_2|^2)/2 + \alpha_1^*\alpha_2}\,,
\end{equation}
we obtain the following simple equation for the overlapping of the initial and displaced SC states:
\begin{equation}\label{overlap}
	\braket{\Psi_0}{\Psi_\delta}
	= \frac{2e^{-\delta^2/2}}{K}(\cos2\alpha\delta + e^{-2\alpha^2}) \,.
\end{equation}
It is plotted in Fig.\,\ref{fig:plot-overlap-by-phi-fixed-alpha-2} as a function of the displacement parameter $\delta$ for two realistic values of the SC amplitudes, \(\alpha=1.5\) and \(\alpha=3.0\).

Note that the wave function $\ket{\Psi_0}$ also corresponds to the output state of light for $\delta=0$ (no phase signal). Therefore, the values of $\phi$ that cancel $\braket{\Psi_0}{\Psi_\delta}$ can be detected unambiguously \cite{HelstromBook}.

It is easy to see that zeros of the function \eqref{overlap} are equal to
\begin{equation}\label{delta_k}
	\delta_k = \frac{\arccos(-e^{-2 \alpha^2}) + 2 \pi k}{2 \alpha} \,,
\end{equation}
where \(k\) is an integer number. Evidently, the best sensitivity is provided by \(k=0\). The corresponding phase shift is equal to
\begin{equation}\label{phi_optimal}
	\phi_0 = \frac{\delta_0}{\sqrt{N}} = \frac{\arccos(-e^{-2 \alpha^2})}{2\alpha\sqrt{N}} \,.
\end{equation}

As $\alpha$ increases, the numerator of \eqref{phi_optimal} quickly converges to $\pi/2$; in particular, if \(\alpha = 1.5\), then the difference is about 1\%. Therefore, $\phi_0$ can be approximated as follows:
\begin{equation}\label{phi_opt_approx}
	\phi_0 \approx \frac{\pi}{4\alpha\sqrt{N}} \,.
\end{equation}

Note the similarity between Eqs.\,\eqref{phi_optimal} and \eqref{dphi_sqz}. In both cases, the right-hand side is inversely proportional to the square root of the carrier photons number. The additional factor \(\arccos(-e^{-2 \alpha^2})/\alpha\) plays the role of the squeeze factor \(e^{-2r}\), allowing to further improve the sensitivity.

At the same time, it has to be emphasized, that these quantities have different meanings: $\Delta\phi_{\rm SQZ}$ defines the mean squared value of the measurement error, while $\phi_0$ is equal to the phase shift that can, in principle, be unambiguously detected.

To achieve this result, the optimal measurement procedure described by the following positive operator-valued measure (POVM):
\begin{equation}\label{POVM_cat}
	\{\ket{\Psi_0}\bra{\Psi_0}, \ket{\Psi_\delta}\bra{\Psi_\delta}\}
\end{equation}
must be used \cite{HelstromBook} (note that this POVM is complete in the two-dimensional space spanned by the states $\ket{\Psi_0}$ and $\ket{\Psi_\delta}$).  Unfortunately, this procedure does not correspond to any of the standard photodetection schemes.

\section{Photon number measurement}\label{sec:detection}

Let us consider a more practical detection procedure based on the measurement of photons number at the dark output port using a photon-number resolving photodetector.

The corresponding photon number statistics is calculated in App.\,\ref{app:p_even_odd}, see Eq.\,\eqref{p_n}. It follows from this equation that the initial SC state $\ket{\Psi_0}$ is a superposition of even Fock states only. In the displaced SC state case, $\delta\neq 0$, the odd Fock states appear and, with the increase of $\delta$, become dominant. Therefore, detection of an odd number of photons guarantees the presence of the phase shift.

In this case, it is natural to use the following strategy: detection of an even number of photons means the decision that $\delta=0$, and of an odd number --- that $\delta\neq0$.

Introduce the $2\times2$ matrix of conditional probabilities $p(\cdot/\cdot)$ of obtaining two possible outcomes, with the second arguments corresponding to the real presence ($\delta\ne0$) or absence of the signal and first one -- to the resulting estimate. We denote the presence and the absence of the signal by ``+'' and ``$-$'', respectively.

The case of $\delta=0$ always gives the ``negative'' result; therefore,
\begin{equation}
  p(-/-) = 1 \,, \quad p(+/-) = 0 \,.
\end{equation}

If $\delta\neq0$, then the ``negative'' and the ``positive'' probabilities are equal to
\begin{equation}
  p(-/+) = p_{\rm even} \,, \quad p(+/+) = p_{\rm odd} \,.
\end{equation}
Here $p_{\rm even}$ and $p_{\rm odd}$ are, respectively, the probabilities of obtaining the even and the odd photon numbers as the result of the measurement.

Note that the error probabilities $p(+/-)$ and $p(-/+)$ are known as the ``false positive'' (or  ``false detection'') and ``false negative'' (or miss signal) ones.

\begin{figure}
	\centering
	\includegraphics[width=1\linewidth]{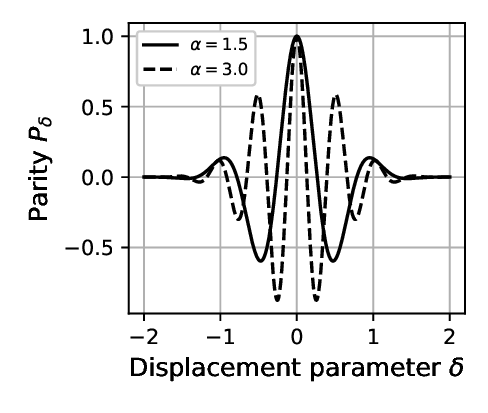}
	\caption{Parity of the displaced SC state, see Eq.\,\ref{parity}) as a function of the displacement parameter \(\delta\) for \(\alpha = 1.5\) (solid line) and \(\alpha = 3\) (dashed line).}
	\label{fig:par-by-delta-alpha}
\end{figure}

The probabilities $p_{\rm even}$ and $p_{\rm odd}$ are calculated in App.\,\ref{app:p_even_odd}, see Eqs.\,\eqref{p_even}, \eqref{p_odd}. They can be presented as:
\begin{equation}\label{p_even_odd}
	  p_{\rm even} = \frac{1 + P_\delta}{2}\,,\quad p_{\rm odd} = \frac{1 - P_\delta}{2} \,,
\end{equation}
where
\begin{equation}\label{parity}
	P_\delta = \bra{\Psi_\delta}\hat{\Pi}\ket{\Psi_\delta}
	= e^{-2\delta^2}\frac{\cos4\alpha\delta + e^{-2\alpha^2}}{1 + e^{-2\alpha^2}}
\end{equation}
is the parity of the state $\ket{\Psi_\delta}$,
\begin{equation}
  \hat{\Pi} = (-1)^{\hat{n}}
\end{equation}
is the parity operator \cite{Birrittella_AVSQS_3_014701_2021}, and $\hat{n}$ is the photon number operator. In Fig.\,\ref{fig:par-by-delta-alpha}, the parity $P_\delta$ is plotted as a function of \(\delta\) for the same values of \(\alpha\) as in Fig.\,\ref{fig:plot-overlap-by-phi-fixed-alpha-2}.

The minimum of the ``false negative'' probability coincides with the minimum of the parity $P_\delta$. The approximate value of $\delta$ that provides this minimum is found in  App.\,\ref{app:min_P}:
\begin{equation}\label{delta_app}
  \delta \approx \frac{\pi}{4\alpha}\frac{1}{1 + \frac{1}{4 \alpha^2}} \,.
\end{equation}
For the values of $\alpha\ge1.25$, this approximation deviates form the exact numerical solution by less that 1\%.

\begin{figure}
	\centering
	\includegraphics[width=1\linewidth]{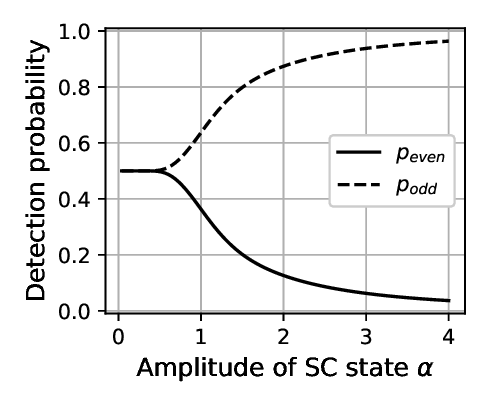}
	\caption{Dependencies of probabilities \(p_{\rm odd}\) and \(p_{\rm even}\) (see Eq.\,\ref{p_even_odd}) on SC state amplitude $\alpha$. Displacement $\delta$ corresponds to the minimum parity of displaced SC state.}
	\label{fig:even-and-odd-probability-ideal-detector}
\end{figure}

In Fig.\,\ref{fig:even-and-odd-probability-ideal-detector}, the probabilities \eqref{p_even_odd} are plotted as functions of \(\alpha\), assuming the optimized values of \(\delta\). It is easy to see for reasonably large values of $\alpha\gtrsim2$, the ``false negative'' probability does not exceed $\sim0.1$.

\section{Results and discussion}\label{sec:conclusion}

We have shown that by injecting the classical (coherent) light into the first (bright) port of a two-arm interferometer and a Shr\"odinger cat state into the second (dark) port, it is possible to unambiguously detect a given phase shift defined by Eq.\,\eqref{phi_optimal}. However, this requires an exotic method of detecting the output light that does not corresponds to any of the ordinary photodetection schemes.

By measuring the number of photons at the output dark port of the interferometer using a photon-number resolving detector, it is still possible to obtain quite interesting results. This procedure allows to detect a given phase shift with a ``false positive'' probability equal to zero. The corresponding ``false negative'' probability in this case decreases monotonically with the increase of the SC amplitude \(\alpha\), see  Fig.\,\ref{fig:even-and-odd-probability-ideal-detector}. For reasonable values of \(\alpha\gtrsim2\), this probability does not exceed $\sim0.1$.

It is worth noting that in many cases, the asymmetric penalty matrices and therefore the unequal values of ``false positive'' and ``false negative'' statistical errors are optimal. The trivial example is the discrimination of poisonous and healthy substances.

Two elements are crucial for the implementation of the proposed scheme: (i) the source of the SC quantum states of light and (ii) the photodetectors which can resolve up to $n=\alpha^2\sim10$ photons in an optical pulse.

The preparation of the SC states with small value of $\alpha^2\approx0.8$ was successfully demonstrated as early as 2006, see Ref.\,\cite{Ourjoumtsev_Science_312_83_2006}. In more recent work, the values of up to \(\alpha^2\sim 3\) were achieved, see e.g. Refs.\,\cite{Huang_PRL_115_023602_2015, Sychev_NPhot_11_379_2017} and the review \cite{Lvovsky2020}. Recently, the methods of preparation of large cat states \(\alpha \gtrsim 4\) \cite{Kuts_D_A_PS_2022}, and \(\alpha \gtrsim 5\)  \cite{Podoshvedov_M_S_SR_2023} with high fidelity using photon number resolving detectors have been proposed.

Regarding the photon number resolving detectors, the superconducting transition edge sensors (TES) can be considered as the best candidate. They can resolve up to \(\sim 10\) photons, and their quantum efficiency can be as high as 98\% \cite{Fukuda_OE_19_870_2011, Gerrits_OE_20_23798_2012, Stasi_L_PRA_19_6_2023}.

Therefore, it is possible to assume that the practical implementation of the scheme discussed in this work can be considered feasible.

\section{Acknowledgments}

This work was supported by the Theoretical Physics and Mathematics Advancement Foundation ``BASIS'' Grant \#23-1-1-39-1.

The author would like to thank F.\,Khalili, B.\,Nugmanov, and D.\,Salykina for the useful remarks and discussions.

\section{Appendices}

\subsection{Derivation of Eq.\,\eqref{ifo_io_H}}\label{app:ifo}

\begin{figure}
	\centering
	\includegraphics[width=1\linewidth]{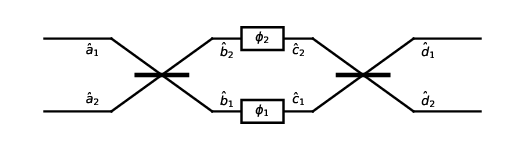}
	\caption The generalized version of the interferometer, considered in App.\,\ref{app:ifo}\label{fig:ifo_app}
\end{figure}

In this Appendix, we consider a generalized optical scheme, see Fig.\,\ref{fig:ifo_app}, that encompasses both asymmetric and antisymmetric options shown in Fig.\,\ref{fig:parity-optic-scheme}(a) and (b). We assume arbitrary real values of the beamsplitters' amplitude reflectivity $R$ and transmissivity $T$, satisfying the unitarity condition $R^2+T^2=1$, and arbitrary phase shifts in the arms $\phi_1$ and $\phi_2$.

Let $\hat{a}_{1,2}$ be the annihilation operators of light at the input ports, $\hat{b}_{1,2}$ --- the ones after the first beamsplitter, $\hat{c}_{1,2}$ --- before the second beamsplitter, and $\hat{d}_{1,2}$ --- at the output ports. In the Heisenberg picture, equations for these operators are the following:
\begin{equation}
	\begin{array}{ll}
		\hat{b}_1 = T\hat{a}_1 - R\hat{a}_2 \,, & \hat{b}_2 = R\hat{a}_1 + T\hat{a}_2 \,,\\
		\hat{c}_1 = \hat{b}_1e^{-i\phi_1} \,, & \hat{c}_2 = \hat{b}_2e^{-i\phi_2} \,, \\
		\hat{d}_1 = T\hat{c}_1 + R\hat{c}_2 \,, & \hat{d}_2 = -R\hat{c}_1 + T\hat{2}_2 \,.
	\end{array}
\end{equation}
Combining these equations, we obtain:
\begin{equation}\label{d1d2}
	\begin{array}{ll}
		\hat{d}_1 = &(T^2e^{-i\phi_1} + R^2e^{-i\phi_2})\hat{a}_1 \\ &
		+ RT(e^{-i\phi_2} - e^{-i\phi_1})\hat{a}_2 \,, \\
		\hat{d}_2 = &RT(e^{-i\phi_2} - e^{-i\phi_1})\hat{a}_1 \\ &
		+ (R^2e^{-i\phi_1} + T^2e^{-i\phi_2})\hat{a}_2 \,.
	\end{array}
\end{equation}

We detach the classical amplitude $A\gg1$ from the quantum fluctuations term at the bright input port:
\begin{equation}
	\hat{a}_1 := A + \hat{a}_1 \,,
\end{equation}
(without loss of generality, we assume that $A$ is a real and positive quantity). Now, the renormalized operator $\hat{a}_1$ corresponds to the vacuum state. We also assume that
\begin{equation}
	|\phi_{1,2}| \ll 1 \,.
\end{equation}
Keeping only linear in $\hat{a}_{1,2}$ and $\phi_{1,2}$ terms in \eqref{d1d2}, we obtain:
\begin{equation}\label{d1d2_app}
	\begin{array}{ll}
		&\hat{d}_1 = A + \hat{a}_1 \,, \\
		&\hat{d}_2 = iRTA(\phi_1 - \phi_2) + \hat{a}_2 \,.
	\end{array}
\end{equation}

Now assume that
\begin{equation}
	\phi_1 = \phi \,, \quad \phi_2 =0 \,, \quad T \to 0 \,, \quad AT = {\rm const}
\end{equation}
for the asymmetric case, and
\begin{equation}
	\phi_1 = -\phi_2 = \phi \,,\quad R = T = \frac{1}{\sqrt{2}} \,.
\end{equation}
--- for the antisymmetric one. In both cases, we come to Eq.\,\eqref{ifo_io_H}.

\subsection{Probabilities of even and odd measured photons}\label{app:p_even_odd}

In the Fock representation, the wave function of the displaced SC state \eqref{Psi_delta} has the following form:
\begin{equation}
	\begin{array}{ll}
		\ket{\psi_{cat}} =& \frac{1}{\sqrt{K}}e^{-(\alpha^2 + \delta^2)/2}
		\sum_n^\infty \frac{1}{\sqrt{n!}} \\[1.5ex]
		&\times
		[e^{i \alpha \delta} (\alpha+i\delta)^n
		+
		e^{-i \alpha \delta} (-\alpha+i\delta)^n]\ket{n} \,.
	\end{array}
\end{equation}
Therefore, the probability distribution for the photons number is equal to
\begin{equation}\label{p_n}
	\begin{array}{ll}
		p_n =& \frac{2}{Kn!}e^{-(\alpha^2 + \beta^2)} \\[1.5ex]
		&\times\Bigl(
		(\alpha^2 + \delta^2)^n + \Re\{e^{2i\alpha\delta}[-(\alpha+i\delta)^2]^n\}
		\Bigr).
	\end{array}
\end{equation}

For the summations over even and odd photon numbers, we use the following equations:
\begin{equation}
	\cosh x = \sum_{n=0}^\infty\frac{x^{2n}}{(2n)!} \,, \quad
	\sinh x = \sum_{n=0}^\infty\frac{x^{2n+1}}{(2n+1)!} \,,
\end{equation}
obtaining:
\begin{equation}\label{p_even}
	\begin{array}{ll}
		p_{\rm even} =& \sum_{n=0}^\infty p_{2n}
		= \frac{2e^{-\alpha^2-\delta^2}}{K}\{
		\cosh(\alpha^2+\delta^2)  \\[1.5ex]
		&+ \Re[e^{2i\alpha\delta}\cosh(\alpha^2-\delta^2+2i\alpha\delta)]
		\} \,,
	\end{array}
\end{equation}
\begin{equation}\label{p_odd}
	\begin{array}{ll}
		p_{\rm odd} =& \sum_{n=0}^\infty p_{2n+1}
		= \frac{2e^{-\alpha^2-\delta^2}}{K}\{
		\sinh(\alpha^2+\delta^2) \\[1.5ex]
		&- \Re[e^{2i\alpha\delta}\sinh(\alpha^2-\delta^2+2i\alpha\delta)]
		\} \,,
	\end{array}
\end{equation}
This result is equivalent to Eqs.\,\eqref{p_even_odd}.

\subsection{The minimum of Eq.\,\eqref{parity}}\label{app:min_P}

The minimum of Eq.\,\eqref{parity} corresponds to the first solution of the following equaiton:
\begin{equation}
  \frac{dP_\delta}{d\delta}
  \propto \delta(\cos4\alpha\delta + e^{-2\alpha^2}) + \alpha\sin4\alpha\delta = 0 \,.
\end{equation}
If the SC state amplitude is sufficiently large, \(e^{-2\alpha^2} \ll 1\), then it can be approximated as follows:
\begin{equation}\label{eq_delta_app}
	\frac{\delta}{\alpha} + \tan(4\alpha\delta) = 0 \,,
\end{equation}
or, equivalently,
\begin{equation}
  \arctan\frac{\delta}{\alpha} + 4\alpha\delta = \pi \,.
\end{equation}
If $\alpha \gg \delta$, then the $\arctan$ can be replaced by its argument, giving the following linear in $\delta$ equation
\begin{equation}
  \frac{\delta}{\alpha} + 4\alpha\delta = \pi \,.
\end{equation}
with the solution equal to \eqref{delta_app}.


\section*{References}

\begin{thebibliography}{100}
	\bibitem{Anderson_ch35_2019} U. L. Andersen, O. Glöckl, T. Gehring, and G. Leuchs
	2019 {\it Quantum interferometry with gaussian states in} {\it Quantum Information} (John Wiley \& Sons, Ltd) Chap. 35 pp. 777-798

	\bibitem{22a1SaKh} D. Salykina and F. Khalili 2023 {\it Symmetry} \textbf{15} 774

	\bibitem{Caves1981} C. M. Caves 1981 {\it Phys. Rev.} D \textbf{23} 1693

	\bibitem{Dwyer_Galaxies10_020046_2022} S. E.Dwyer, G. L. Mansell, and L. McCuller 2022 {\it Galaxies} \textbf{10}

	\bibitem{Ou_PRL_77_2352_1996} Z. Y. Ou 1996 {\it Phys. Rev. Lett.} \textbf{77} 2352

	\bibitem{Ou_PRA_55_2598_1997} Z. Y. Ou 1997 {\it Phys. Rev.} A \textbf{55} 2598

	\bibitem{17a1MaKhCh} M. Manceau, F. Khalili, and M. Chekhova 2017 {\it New Journal of Physics} \textbf{19} 013014

	\bibitem{Schleich2001} W. Schleich 2001 {\it Quantum Optics in Phase Space} (WILEY-VCH, Berlin) p. 695

	\bibitem{Holland_PRL_71_1355_1993} M. J. Holland and K. Burnett 1993 {\it Phys. Rev. Lett.} \textbf{71} 1355

	\bibitem{Lee_JMO_49_2325_2002} H. Lee, P. Kok, and J. P. Dowling 2002 {\it Journal of Modern Optics} \textbf{49} 2325

	\bibitem{Campos_PRA_68_023810_2003} R. A. Campos, C. C. Gerry, and A. Benmoussa 2003 {\it Phys. Rev.} A \textbf{68} 023810

	\bibitem{Berry_PRA_80_052114_2009} D.W. Berry, B. L. Higgins, S. D. Bartlett, M.W. Mitchell, G. J.	Pryde, and H. M. Wiseman 2009 {\it Phys. Rev.} A \textbf{80} 052114

	\bibitem{Pezze_PRL_110_163604_2013} L. Pezzé and A. Smerzi 2013 {\it Phys. Rev. Lett.} \textbf{110} 163604

	\bibitem{Daryanoosh_NComm_9_4606_2018} S. Daryanoosh, S. Slussarenko, D. W. Berry, H. M. Wiseman, and G. J. Pryde 2018 {\it Nature Communications} \textbf{9} 4606

	\bibitem{Perarnau-Llobet_QST_5_025003_2020} M. Perarnau-Llobet, A. González-Tudela, and J. I. Cirac 2020 {\it Quantum Science and Technology} \textbf{5} 025003

	\bibitem{Shukla_OptQEl_55-460_2023} G. Shukla, K. M. Mishra, A. K. Pandey, T. Kumar, H. Pandey, and D. K. Mishra 2023 {\it Optical and Quantum Electronics} \textbf{55} 460

	\bibitem{Shukla_PhOpen_18_100200_2024} G. Shukla, D. Yadav, P. Sharma, A. Kumar, and D. K. Mishra 2024 {\it Physics Open} \textbf{18} 100200

	\bibitem{Lang_PRL_111_173601_2013} M. D. Lang and C. M. Caves 2013 {\it Phys. Rev. Lett.} \textbf{111}, 173601.

	\bibitem{Lang_PRA_90_025802_2014} M. D. Lang and C. M. Caves 2014 {\it Phys. Rev.} A \textbf{90}, 025802

	\bibitem{HelstromBook} C.W. Helstrom 1976 {\it Quantum Detection and Estimation Theory} (Academic Press, New York) p. 309

	\bibitem{Wolf_NComm_10_2929_2019} F. Wolf, C. Shi, J. C. Heip, M. Gessner, L. Pezzè, A. Smerzi,	M. Schulte, K. Hammerer, and P. O. Schmidt 2019 {\it Nature Communications} \textbf{10} 2929

	\bibitem{Singh_PhOpen_18_100198_2024} R. Singh and A. E. Teretenkov 2024 {\it Physics Open} \textbf{18} 100198

	\bibitem{CQG_32_7_074001_2015} J.Aasi {\it et al} 2015 {\it Classical and Quantum Gravity} \textbf{32} 074001

  \bibitem{Birrittella_AVSQS_3_014701_2021} R. J. Birrittella, P. M. Alsing, and C. C. Gerry 2021 {\it AVS Quantum Science} {\bf 3}, 014701,

	\bibitem{Ourjoumtsev_Science_312_83_2006} A. Ourjoumtsev, R. Tualle-Brouri, J. Laurat, and P. Grangier 2006 {\it Science} \textbf{312} 83

	\bibitem{Huang_PRL_115_023602_2015} K. Huang {\it et al} 2015 {\it Phys. Rev. Lett.} \textbf{115} 023602

	\bibitem{Sychev_NPhot_11_379_2017} D. V. Sychev, A. E. Ulanov, A. A. Pushkina, M. W. Richards, I. A. Fedorov, and A. I. Lvovsky 2017 {\it Nature Photonics} \textbf{11} 379

	\bibitem{Lvovsky2020} A. I. Lvovsky, P. Grangier, A. Ourjoumtsev, V. Parigi, M. Sasaki,	and R. Tualle-Brouri 2020 arXiv:2006.16985

	\bibitem{Kuts_D_A_PS_2022} Kuts, D.A., Podoshvedov, M.S., Nguyen, B.A. and Podoshvedov, S.A. 2022 {\it Physica Scripta} \textbf{97}(11) 115002

	\bibitem{Podoshvedov_M_S_SR_2023} Podoshvedov, M.S., Podoshvedov, S.A. and Kulik, S.P. 2023 {\it Scientific Reports} \textbf{13}(1) 3965

	\bibitem{Fukuda_OE_19_870_2011} Fukuda, D. et al 2011 {\it Optics express} \textbf{19}(2) 870

	\bibitem{Gerrits_OE_20_23798_2012} Gerrits, T., Calkins, B., Tomlin, N., Lita, A.E., Migdall, A., Mirin, R. and Nam, S.W. 2012 {\it Optics Express} \textbf{20}(21) 23798

	\bibitem{Stasi_L_PRA_19_6_2023} Stasi, L., Gras, G., Berrazouane, R., Perrenoud, M., Zbinden, H. and Bussières, F. 2023 {\it Physical Review Applied} \textbf{19}(6) 064041

\end{thebibliography}
\end{document}